\documentclass[12pt,letterpaper]{article}
\usepackage{amsmath,amssymb,array,calc,rotating,epsfig,psfrag,amscd, cite}

\newcommand{\nc}{\newcommand}

 \nc{\ra}{\rightarrow} 
\def\al{\alpha}

\nc{\veps}{\varepsilon}
\def\gam{\gamma}

\def\om{\omega}
\nc{\vphi}{\varphi}
\def\tha{\theta}

\def\Gam{\Gamma}

\def\Om{\Omega}
\def\Sig{\Sigma}

\nc{\bea}{\begin{eqnarray}}
\nc{\eea}{\end{eqnarray}}
\nc{\be}{\begin{equation}}
\nc{\ee}{\end{equation}}
\nc{\cA}{{\cal A}}
\nc{\cB}{ \cal B}

\nc{\cF}{{\cal F}}
\nc{\cG}{{\cal G}}

\nc{\cL}{{\cal L}}
\nc{\M}{{\cal M}}
\nc{\cM}{{\cal M}}

\nc{\cQ}{{\cal Q}}
\nc{\cR}{{\cal R}}

\def\T{{\cal T}}

\nc{\BB}{{\mathbb B}}
\nc{\CC}{{\mathbb C}}
\nc{\DD}{{\mathbb D}}
\nc{\EE}{{\mathbb E}}
\nc{\FF}{{\mathbb F}}
\nc{\GG}{{\mathbb G}}
\nc{\HH}{{\mathbb H}}
\nc{\JJ}{{\mathbb J}}
\nc{\RR}{{\mathbb R}}
\nc{\PP}{{\mathbb P}}
\nc{\QQ}{{\mathbb Q}}
\nc{\ZZ}{{\mathbb Z}}
\nc{\CP}{{\CC\PP}}
\nc{\calone}{{\mathbb 1}}
\nc{\half}{\frac{1}{2}}
\nc{\qrt}{\frac{1}{4}}
\nc{\del}{\partial}

\nc{\delbar}{\bar\partial}
\nc{\Spin}{\operatorname{Spin}}
\nc{\SO}{\operatorname{SO}}

\nc{\Sp}{{\rm Sp}}
\nc{\com}[2]{{ \left[ #1, #2 \right] }}
\nc{\acom}[2]{{ \left\{ #1, #2 \right\} }}
\nc{\rr}{\rightarrow}
\nc{\p}{\partial}
\nc{\LT}{{\LL_\T}}
\nc{\Tr}{{\rm Tr}}
\nc{\tr}{{\rm tr}}
\def\com#1#2{{ \left[ #1, #2 \right] }}
\def\acom#1#2{{ \left\{ #1, #2 \right\} }}
\nc{\tKT}{\widetilde{K3}}
\nc{\ttha}{\tilde{\theta}}
\nc{\tphi}{\tilde{\phi}}
\nc{\tPhi}{\tilde{\Phi}}
\nc{\tpsi}{\tilde{\psi}}
\nc{\tgam}{\tilde{\gam}}
\nc{\tGam}{\tilde{\Gam}}
\nc{\tSig}{\tilde{\Sig}}
\nc{\tc}{\tilde c}
\nc{\te}{\tilde e}
\nc{\tg}{\tilde g}
\nc{\tj}{\tilde j}
\nc{\tp}{\widetilde{p}}
\nc{\tq}{\widetilde{q}}
\nc{\ts}{{\tilde s}}
\nc{\tz}{\tilde z}
\nc{\tD}{{\tilde D}}
\nc{\tE}{{\tilde E}}
\nc{\tG}{{\tilde G}}
\nc{\tH}{{\tilde H}}
\nc{\tM}{{\tilde M}}
\nc{\tN}{{\tilde N}}
\nc{\tP}{{\tilde P}}
\nc{\tQ}{{\tilde Q}}
\nc{\tS}{\tilde{S}}
\nc{\tF}{\tilde{{\cal F}}}
\nc{\tX}{\widetilde{X}}
\nc{\hb}{\hat b}
\nc{\hc}{\hat c}
\nc{\hd}{\hat d}
\nc{\he}{\hat e}
\nc{\hf}{\hat f}
\nc{\hg}{\hat g}
\nc{\hh}{\hat h}
\nc{\hp}{\hat p}
\nc{\hw}{\hat w}
\nc{\hx}{\hat x}
\nc{\hy}{\hat y}
\nc{\hz}{\hat z}
\nc{\hA}{\widehat{A}}
\nc{\hE}{\widehat{E}}
\nc{\hH}{\widehat{H}}
\nc{\hJ}{\widehat{J}}
\nc{\tK}{\widetilde{K}}
\nc{\hM}{\widehat M}
\nc{\hF}{\widehat{\F}}

\nc{\ha}{\widehat \alpha}
\nc{\hphi}{\hat{\phi}}
\nc{\hpsi}{\hat{\psi}}
\nc{\hgam}{\hat{\gam}}
\nc{\hPhi}{\hat{\Phi}}
\nc{\hPsi}{\hat{\Psi}}
\nc{\hGam}{\hat{\Gam}}

\nc{\w}{\wedge}

\nc{\ol}{\overline}
\nc{\abar}{\ol{a}}
\nc{\bbar}{\ol{b}}
\nc{\cbar}{\ol{c}}
\nc{\ebar}{\ol{e}}
\nc{\ibar}{\ol{\imath}}
\nc{\jbar}{\ol{\jmath}}
\nc{\kbar}{\ol{k}}
\nc{\lbar}{\ol{l}}
\nc{\mbar}{\ol{m}}
\nc{\nbar}{\ol{n}}
\nc{\ubar}{\ol{u}}
\nc{\vbar}{\ol{v}}
\nc{\wbar}{\ol{w}}
\nc{\xbar}{\ol{x}}
\nc{\ybar}{\ol{y}}
\nc{\zbar}{\ol{z}}

\nc{\Ebar}{\ol{E}}
\nc{\Jbar}{\ol{J}}
\nc{\Qbar}{\ol{Q}}
\nc{\Wbar}{\ol{W}}
\nc{\Xbar}{{\overline X}}
\nc{\Ybar}{{\overline Y}}
\nc{\Zbar}{{\overline Z}}

\nc{\epsbar}{\ol{\epsilon}}
\nc{\lambar}{\ol{\lambda}}
\nc{\psibar}{\ol{\psi}}
\nc{\Psibar}{\ol{\Psi}}
\nc{\phibar}{\ol{\phi}}
\nc{\Phibar}{\ol{\Phi}}
\nc{\chibar}{\ol{\chi}}
\nc{\ombar}{\ol{\om}}
\nc{\Ombar}{\ol{\Om}}
\nc{\bah}{{\mathbf {\hat{A}}}}
\nc{\bX}{{\mathbf X}}
\nc{\dal}{\dot{\al}}
\nc{\thab}{\bar{\theta}}
\nc{\thal}{\theta^{\al}}
\nc{\thdal}{\bar{\theta}^{\dal}}

\nc{\thsigthm}{\tha \sigma^m \thab}
\nc{\thsigthn}{\tha \sigma^n \thab}

\nc{\Dal}{D_{\al}}
\nc{\Ddal}{\bar{D}_{\dal}}
\nc{\CDal}{{\cal D}_{\al}}
\nc{\CDdal}{\bar{\cal D}_{\dal}}

\nc{\eq}[1]{(\ref{#1})}
\nc{\non}{\nonumber}
\nc{\comment}[1]{{\bf #1}}
\nc{\xs}{\not\!\!X}
\nc{\ps}{\not\!\!P}
\nc{\dif}{{d}}
\nc{\equ}{{\rm eq}}

\nc{\AdS}{{\rm AdS}}
\nc{\vol}{{\rm vol}}
\nc{\Ainf}{A_{\infty}}
\nc{\End}{{\rm End}}
\nc{\Ext}{{\rm Ext}}
\nc{\Hom}{{\rm Hom}}
\nc{\IIB}{{\rm IIB}}
\nc{\Pic}{{\rm Pic}}
\nc{\bra}[1]{\langle{#1}|}
\nc{\ket}[1]{|{#1}\rangle}
\nc{\braket}[2]{\langle{#1}|{#2}\rangle}
\nc{\sect}[1]{Section~\ref{#1}}
\nc{\fig}[1]{Fig.~\ref{#1}}
\nc{\chap}[1]{Chapter~\ref{#1}}
\nc{\Dslash}{\ensuremath \raisebox{0.025cm}{\slash}\hspace{-0.32cm} D}
\nc{\no}{\!:\!\!}
\nc{\bpm}{\begin{pmatrix}}
\nc{\epm}{\end{pmatrix}}
 \nc{\bitem}{\begin{itemize}}
 \nc{\eitem}{\end{itemize}}

\newcommand{\C}[1]{$(\ref{#1})$}

\def\Z{\mathbb{Z}}

\def\SO{\operatorname{SO}}

\nc{\rank}{{\rm rank}} 
\nc{\pr}{{\rm pr}} 

\nc{\tom}{\tilde{\om}} 
\nc{\tOm}{\tilde{\Om}}

\begin{document}

\begin{titlepage}

\begin{center}


{July 2, 2007} \hfill     EFI-07-19
\vskip 2 cm
{\Large \bf A Note on Heterotic Dualities via M-theory}\\
\vskip 1.25 cm 
Savdeep Sethi\footnote{sethi@theory.uchicago.edu} \\

{ \vskip 0.5cm Enrico Fermi Institute, \\
University of Chicago, \\Chicago, IL 60637, USA\\}

\end{center}

\vskip 0.5 cm

\begin{abstract}
\baselineskip=18pt

We show that a class of torsional compactifications of the heterotic string are dual to conventional K\"ahler heterotic string compactifications. This observation follows from the recently proposed analogue of the c-map for the
heterotic string.

\end{abstract}

\end{titlepage}


\baselineskip=18pt

Within the past few years, there has been significant improvements in our understanding of string vacua. Most of the recent developments involve turning on background fluxes starting with~\cite{Becker:1996gj}. Compactifications with flux tend to be difficult to analyze: either the background involves RR fluxes which complicates a world-sheet analysis or the backgrounds are intrinsically quantum with cycles of order the string scale. In special cases, however, there are dualities that relate flux compactifications to more conventional geometric compactifications. Dualities of this kind appear in~\cite{Sethi:1996es, Berglund:2005dm, Schulz:2004tt}. In these cases, we can use the powerful techniques available for conventional compactifications to learn about backgrounds with flux. 
  
The aim of this note is to present a duality of this kind that relates a class of standard compactifications of the heterotic string on K\"ahler  spaces to compactifications of the heterotic string on non-K\"ahler spaces with torsion. The torsion corresponds to a non-trivial background $H_3$-flux. In the past, these torsional compactifications were studied using supergravity~\cite{Strominger:1986uh}\ and a perturbative sigma model analysis~\cite{Gates:1984nk, Hull:1986kz}. Compact torsional solutions are quite difficult to construct. The class of known compact solutions take the form of torus bundle twisted over a $K3$ surface with metric $g_{K3}$. The solutions have the schematic form
\be\label{DRS}
ds^2 =  \Delta g_{K3} + (d\theta^1 + A_1)^2 + (d\theta^2+ A_2)^2 
\ee
where $\Delta$ is a warp factor and $(\theta^1,\theta^2)$ parametrize $T^2$. The connections $(A_1, A_2)$ are $1$-forms on the base $K3$ surface which determine the twisting of the torus fiber. Correlated with this twist is an $H_3$ flux.
These are the DRS torsional solutions~\cite{Dasgupta:1999ss}. 

There has been much subsequent development of torsional backgrounds; a partial list of references includes~\cite{Becker:2002sx, LopesCardoso:2002hd, Goldstein:2002pg, LopesCardoso:2003af,  Becker:2003yv, Gauntlett:2003cy, Becker:2003gq, Becker:2003sh, LopesCardoso:2003sp, Schulz:2004ub,  Becker:2005nb,  Fu:2006vj, Becker:2006et, Dasgupta:2006yd, Kimura:2006af, Knauf:2006mz, Kim:2006qs, Becker:2006xp}. It is important to note that while the metric and fluxes locally satisfy the conditions described in~\cite{Strominger:1986uh, Gates:1984nk, Hull:1986kz}, the global solution involves cycles of $O(\alpha')$. The solutions are therefore inherently stringy. 

The duality we will present relates the DRS torsional solutions~\C{DRS}\ and mild generalizations to conventional heterotic backgrounds on K\"ahler spaces. This duality is a straightforward corollary of the c-map~\cite{Cecotti:1988qn}\ for the heterotic string proposed recently in~\cite{Halmagyi:2007wi}.\footnote{ For another approach relating the DRS torsional metrics to standard heterotic compactifications, see~\cite{Adams:2007vp}.}  

Our starting point is the particularly nice example of M-theory on $K3\times K3'$ constructed in~\cite{Dasgupta:1999ss}. Let us denote the first $K3$ surface by $S$ and the second by $S'$. This background has a net non-vanishing  M2-brane charge which must be canceled~\cite{Becker:1996gj, Sethi:1996es}  The cancelation can be accomplished by a combination of  $n_{M_2}$ inserted M2-branes and $4$-form flux $G_4$ satisfying the tadpole constraint
\be\label{Mtadpole}
{1\over 2} \int{ {G_4 \over 2\pi} \wedge {G_4 \over 2\pi}} + n_{M_2} = 24. 
\ee
We will want to restrict to $n_{M_2}=0$ if we wish to find a standard perturbative heterotic dual. We must therefore turn on $G_4$ flux which satisfies the requirements
\be {G_4\over 2\pi} \in H^{2,2}(S \times S', \Z) \ee 
and that $G_4$ be primitive. Satisfying these requirements fixes some of the complex and K\"ahler moduli of the compactification. Prior to turning on the flux, this compactification preserves N=4 supersymmetry in three dimensions. If the choice of $G_4$ flux is $(2,2)$ and primitive with respect to each of the $\CP^1 \times \CP^1$ choices of complex structure then the full N=4 is preserved. Otherwise, the flux will preserve only N=2 supersymmetry; examples of both kinds can be found in~\cite{Dasgupta:1999ss}.  

The flux can be expressed as follows, 
\be\label{Gflux}
{G_4 \over 2\pi} = \omega \wedge \omega', 
\ee 
where $\omega \in H^2(S, \Z)$ while $\omega' \in H^2(S', \Z)$. If $\omega$ and $\omega'$ are purely of $(1,1)$ type then the flux preserves the full N=4 supersymmetry. This flux compactification will serve as a bridge relating two heterotic compactifications. 

To obtain the first heterotic compactification, we will use the duality between M-theory on a $K3$ surface and the heterotic string on $T^3$~\cite{Witten:1995ex}. Let us assume that both $S$ and $S'$ are elliptically-fibered $K3$ surfaces with section. The choice of an elliptic fibration corresponds to the choice of a circle in $T^3$ for the heterotic dual. This assumption is unnecessary if we only wish to consider three-dimensional heterotic compactifications. It plays a role only if we want to discuss four-dimensional compactifications.  

Let us denote the volume of $S$ (or $S'$) by $V$ (or $V'$). We will measure all our volumes in eleven-dimensional Planck units with $\ell_p=1$ for simplicity. The volume of the elliptic fiber is denoted by $E$ (or $E'$) and the volume of the base by $B$ (or $B'$). The three-dimensional duality equates
\begin{eqnarray} \label{mapone}
\text{M-theory on}~S\times S' &\leftrightarrow& \text{Het. on}~S_H \times T^2\times S^1_R , \nonumber\\
  (E, V,E',V')        &\leftrightarrow&    (E_H, V_H, \lambda_3, R ). 
\end{eqnarray}
The heterotic string compactification is characterized by the three-dimensional string coupling $\lambda_3$, the size $R$ of the distinguished circle $S^1_R $ and volume $V_H$ of the $K3$ surface $S_H$ with elliptic fiber $E_H$. 
These parameters are related in the following way~\cite{Halmagyi:2007wi}:
\begin{eqnarray}
E_H & = & V' E, \nonumber\\
V_H & = & {V'}^2 V, \nonumber\\
R        & = & {V'}^{1/2} E'^{-1}, \nonumber\\
\lambda_3 & = & {V'}^{-1/4} V^{-1/2}.
\end{eqnarray}
Taking $R\rightarrow \infty$ while holding fixed the four-dimensional heterotic coupling $\lambda_4 = \sqrt{R} \lambda_3$ corresponds to taking $E' \rightarrow 0$. This is the F-theory limit~\cite{Vafa:1996xn}. The result is a four-dimensional heterotic compactification on one side of the equivalence and a type IIB flux compactification on the other~\cite{Dasgupta:1999ss}. 

In addition to the parameter map, we need to specify the gauge bundle in the heterotic string. This works beautifully in three dimensions. At generic points in the moduli space, the heterotic string on $T^3$ has 22 abelian gauge-fields. These gauge-fields arise in M-theory by reducing the $C_3$-form potential on the $22$ elements of $H^2(K3, \Z)$.
So we can view the heterotic gauge-fields $A$ as arising from the $G_4$ flux of M-theory via the reduction
\be
{C_3 \over 2\pi} = A \wedge \omega' 
\ee 
with field-strength
\be\label{hetbundle}
{G_4 \over 2\pi} = F_2 \wedge \omega'  = \omega \wedge \omega'. 
\ee
Now the lattice  $H^2(K3, \Z)$ is the even self-dual lattice $\Gamma^{3,19}$ with signature $(3,19)$ and decomposition
\be
\Gamma^{3,19} =  3\Gamma^{1,1}\oplus 2 (-\Gamma^{E_8}). 
\ee
In terms of this decomposition, we can associate gauge-fields coming from reducing on $(-\Gamma^{E_8}) \oplus (-\Gamma^{E_8})$ to the ten-dimensional gauge-fields of the heterotic string. Those coming from reducing $C_3$ on $3\Gamma^{1,1}$ can be viewed as Kaluza-Klein gauge-fields arising in the heterotic string from reducing the $B$-field and metric, $g$, of supergravity on $T^3$.

Since we chose to use no M2-branes to cancel the charge tadpole, the dual heterotic compactification possesses no NS5-branes. For a compactification on $K3 \times T^3$, heterotic anomaly cancelation requires a gauge bundle with instanton number $24$. The $G_4$ flux provides precisely such a bundle when the tadpole condition~\C{Mtadpole} is satisfied. 

So far we have not said anything particularly new. To see something interesting, let us focus on the gauge-fields that arise from $3\Gamma^{1,1}$. If we chose a $G_4$ flux with components in these directions then in the dual heterotic string, we are giving a field strength over the $K3$ surface to the Kaluza-Klein gauge-fields
\be
g_{\mu i} d\theta^i, \quad B_{\mu i} d\theta^i
\ee
where $\theta^i$ coordinatize the $T^3$ factor. This corresponds to twisting the metric of the $T^3$ factor over the $K3$ surface $S_H$. Concomitant with this twist is an $H_3$ flux. This is precisely the structure of the DRS torsional solution~\cite{Dasgupta:1999ss}\ but we have avoided using the original duality chain to see that the solution exists.  For special choices of $G_4$ flux which admit an F-theory limit, we could repeat the steps of~\cite{Dasgupta:1999ss}\ to find the dual torsional compactification. 

More general choices of $G_4$ flux and more general $K3$ metrics can obstruct the F-theory limit. For such choices, the dual heterotic theory is honestly $T^3$-fibered. There is no distinguished $S^1_R$ factor which we can decompactify to obtain a four-dimensional theory.    

Now following~\cite{Aspinwall:2005qw, Halmagyi:2007wi}, we can exchange the roles of $S$ and $S'$ in M-theory. We use the same map as in~\C{mapone}\ to find a second heterotic dual. The key point (as in~\cite{ Halmagyi:2007wi}) is that the roles of $\omega$ and $\omega'$ are exchanged. The heterotic gauge bundle in this second compactification has a field-strength proportional to $\omega'$ rather than $\omega$, 
\be\label{hetbundle}
{G_4 \over 2\pi} = \omega \wedge F_2  = \omega \wedge \omega'. 
\ee
If we choose $\omega$ to lie in the $(-\Gamma^{E_8}) \oplus (-\Gamma^{E_8})$ component of $H^2(K3', \Z)$ then this compactification will involve no torsion. The heterotic string target space will be the product manifold $K3'\times T^3$.  
In the case of an N=4 compactification, this exchange of $S$ and $S'$ provides an analogue of the c-map for the heterotic string, exchanging hypermultiplets and vector multiplets~\cite{Halmagyi:2007wi}. The main novelty we are adding here is the observation that the map extends to N=2 compactifications and further relates torsional compactifications to conventional heterotic string compactifications.   

There are a few additional points worth noting. The heterotic strings in these constructions emerge in M-theory by wrapping an M5-brane on either $S$ or $S'$.  The KK reduction of the M5-brane on a $K3$ surface without flux has been studied in~\cite{Cherkis:1997bx}\ and for more general wrappings~\cite{Minasian:1999qn}. The extension that includes $G_4$ flux is going to result in a string with a quite interesting sigma model.  The $G_4$ flux couplings to fermions on the M5-brane has been investigated in~\cite{Saulina:2005ve, Kallosh:2005yu}.

Imagine wrapping an M5-brane on $S'$ of~\C{mapone}. If we choose $G_4$ flux of the form~\C{Gflux}\ and choose $\omega'$ to lie in the $(-\Gamma^{E_8}) \oplus (-\Gamma^{E_8})$ component of $H^2(S', \Z)$ then we expect a $(0,4)$ heterotic string with standard K\"ahler target space.  For a more general choice of $G_4$, we expect $(0,2)$ world-sheet supersymmetry and a torsional target space. Both possibilities are unified in the reduction of this M5-brane.  This construction has the advantage of being easily generalized. If replace $S \times S'$ by a more general CY $4$-fold with flux, reducing the M5-brane on a supersymmetric $4$-cycle will typically give a $(0,2)$ heterotic string with torsion.  Even simple cases like Nikulin quotients~\cite{Nikulin}\ of $S\times S'$ should result in nice generalizations of~\cite{Dasgupta:1999ss}. 

Finally, we can ask whether this three-dimensional duality between heterotic strings can be pushed to four dimensions. For example, taking $E' \rightarrow 0$ in~\C{mapone}\ nicely results in a four-dimensional heterotic dual, assuming the $G_4$ flux does not obstruct the F-theory limit.  After exchanging the roles of $S$ and $S'$ to obtain a second heterotic theory on $(T^3)' \times S_H'$, we find that the four-dimensional limit corresponds to shrinking the elliptic fiber of the heterotic $K3$ surface $S_H'$. This highly quantum limit of the heterotic string was described in~\cite{Halmagyi:2007wi}: the emergence of a new dimension comes about from light wrapping modes of NS5-branes on $(T^3)' \times E_H'$. It would be interesting to see the extent to which the $K3$ conformal field theory knows about the emergence of this new dimension.

\vskip 0.5 cm

{\bf \noindent Note added:} During the completion of this work, a paper appeared with overlapping results~\cite{Becker:2007ea}.

\section*{Acknowledgements}
It is my pleasure to thank I.~Melnikov and E.~Martinec for helpful discussions. The work S.~S. is supported in part by NSF CAREER Grant No. PHY-0094328 and by NSF Grant No. PHY-0401814. 


\providecommand{\href}[2]{#2}\begingroup\raggedright\endgroup

\end{document}